\documentclass[a4paper,11pt,notitlepage]{article}
\usepackage{multirow,url,color}
\usepackage[utf8x]{inputenc}

\title{\textbf{NIEL DOSE and DLTS Analyses on Triple and Single Junction solar cells irradiated with electrons and protons}}
\author{Roberta Campesato$^{1}$, Carsten Baur$^{5}$, Mariacristina Casale$^{1}$, Massimo Gervasi$^{2,3}$,\\ Enos Gombia$^{4}$, Erminio Greco$^{1}$, Aldo Kingma$^{4}$, Pier Giorgio  Rancoita$^{2}$,\\ Davide Rozza$^{2,3}$, Mauro Tacconi$^{2,3}$.}
\date{}

\usepackage[utf8x]{inputenc}
\usepackage{amsmath}
\usepackage{amsfonts}
\usepackage{amssymb}
\usepackage{graphicx}
\usepackage[top=2.3cm,bottom=2.3cm,left=2.3cm,right=2.3cm]{geometry}

\begin{document}
\maketitle
\begin{center}
$^1$ \textit{CESI, via Rubattino 54, I-20134 Milan, Italy}\\
$^2$ \textit{INFN Sezione di Milano Bicocca, I-20126 Milan, Italy}\\
$^3$ \textit{Universit\`a di Milano Bicocca, I-20126 Milan, Italy} \\
$^4$ \textit{IMEM-CNR Institute, Parco Area delle Scienze 37/A, 43124 Parma, Italy}\\
$^5$ \textit{ESA/ ESTEC, Keplerlaan 1, 2201 AZ Noordwijk, The Netherlands}\\
\vspace{0.5cm}
to appear in the Proceedings of the 
\par 
World Conference on Photovoltaic Energy Conversion (WCPEC-7),\par  Waikoloa, HAWAII, June 10-15, 2018.
\end{center}

\begin{abstract}
Space solar cell radiation hardness is of fundamental importance in view of the future missions towards harsh radiation environment (like the Jupiter missions) and for the new spacecraft using Electrical Propulsion. In this paper we report the radiation data for triple junction (TJ) solar cells and related component cells. Triple junction solar cells, InGaP top cells and GaAs middle cells degrade after electron radiation as expected. With proton irradiation, a high spread in the remaining factors was observed, especially for the TJ and Ge bottom cells. Radiation results have been analyzed by means of the Displacement Damage Dose method and DLTS spectroscopy. In particular with DLTS spectroscopy it was possible to analyze the nature of a few defects introduced by irradiation  inside  the GaAs sub cell observing a strong correlation with the Displacement Damage Dose.\\
\par Index Terms - Photovoltaic cells.
\end{abstract}

\section{Introduction}
In the last 10 years spacecraft are mainly powered by 30$\%$  efficient Triple junction solar cells based on III-V compound semiconductors. The radiation analysis of solar cells is very important to predict the End Of Life (EOL) performances of the solar arrays.
To understand the nature of the defects introduced by irradiation, a powerful technique is the DLTS technique (Deep level Transient Spectroscopy).
DLTS is applicable to single junction devices and up to now ad hoc diodes based on InGaP and GaAs were manufactured to study their behavior after irradiation using such method.
This is the first time that the InGaP and InGaAs samples used for DLTS have exactly the same epitaxial structure of the sub-cell composing the triple junction device. Of course, the usage of DLTS is more difficult because, for example, of higher doping concentrations employed for these samples with respect to a dedicated test diode.
The test of the solar cell radiation hardness is conducted on Earth by irradiating the solar cells using protons and electrons at different energies.
The evaluation of the radiation hardness of the solar cells is performed by means of two methods: The Equivalent Fluence method from JPL \cite{Anspaugh} or The Displacement Damage Dose (DDD) from NRL \cite{Messengers}.
In this paper, we will present the results of electron and proton irradiation on triple junction solar cells and related component cells manufactured by CESI and the radiation hardness analysis conducted by the DDD method.
Furthermore, as already mentioned, the top and middle samples were manufactured as diodes for DLTS measurements and irradiated together with the solar cells.

\section{Triple Junction Solar Cells And Component Cells}
\begin{figure}[h!]
\centering
	\includegraphics[width=0.8\textwidth]{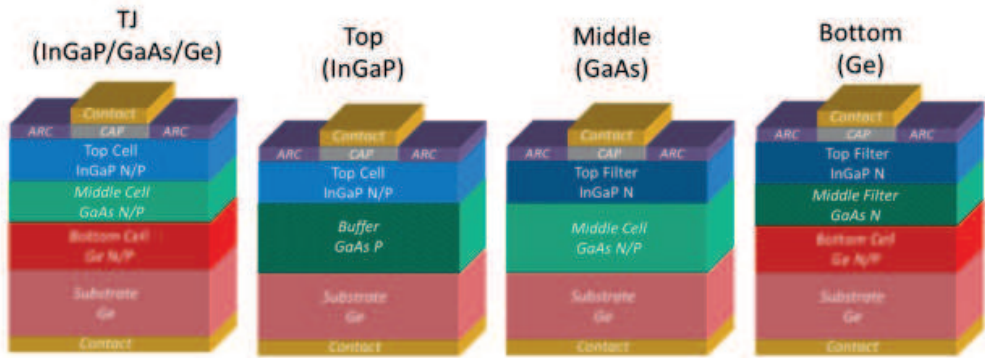}
\caption{Scheme of a TJ cell, top, mid, bot.}
\label{fig:schema}
\end{figure}

\begin{figure}[h!]
\centering
\includegraphics[width=0.5\textwidth]{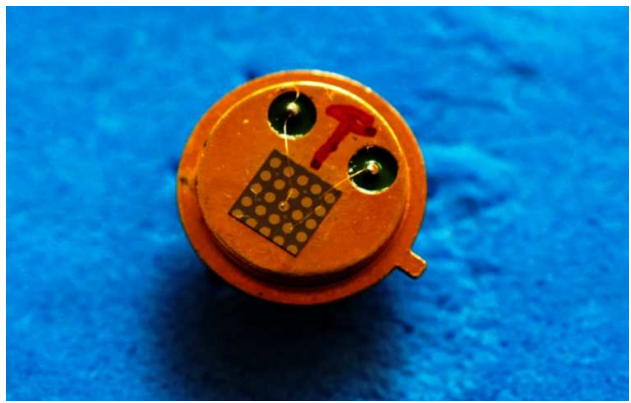}
\caption{0.5 mm dia diodes for DLTS and irradiation analysis.}
\label{fig:2}
\end{figure}

InGaP/InGaAs/Ge TJ solar cells and related component cells  with AM0 efficiency class 30$\%$  (CTJ30), have been manufactured as $2\times2$ cm$^2$ solar cells and 0.5 mm dia diodes (only top and middle sub cell) \cite{Gori}.
The basic structure of the solar cells is reported in Fig. \ref{fig:schema}. The TJ solar cell is composed by a germanium bottom junction obtained by diffusion into the germanium P-type substrate, a middle junction of (In)GaAs, whose energy gap is around 1.38 eV and a top junction of InGaP with an energy gap of 1.85 eV. Component cells are single-junction cells which shall be an electrical and optical representation of the subcells inside the TJ cell. Therefore, to manufacture them, special attention was put to reproduce the optical thicknesses of all the upper layers present in the TJ structure.
Top and mid sub cells were also manufactured as diodes with 0.5 mm diameter using mesa etch to remove the edge defects related to cutting. Fig. \ref{fig:2} shows a picture of a diodes rack used for DLTS analysis.

\section{Experimental Irradiation Results}
TJ solar cells and  component cells have been irradiated with protons and electrons\  at different energies and fluencies \cite{Campesato}.
Solar cells have been measured in BOL conditions and then after irradiation and after the annealing consisting, as per ECSS standards\  in 48 hours at 1 AM0 illumination followed by 48 h at 60$ ^{\circ} $ C. The main electrical parameters were recorded and the ratio EOL/BOL (known as Remaining Factor) was calculated. The bottom junction is highly degraded after proton irradiation whereas it is highly radiation resistant when irradiated with electrons.
After annealing, the Voc of top and mid cells increases thus improving the Voc of the TJ as expected. The bottom junction seems to recover a portion of the short circuit current (+10$\%$  after annealing) but the shunted I-V curve is still present.
The analysis of remaining factors against DDD is reported in the next chapter.

\section{NIEL Analysis}
The photovoltaic parameters of the TJ solar cell and single junction cells are investigated as a function of displacement damage dose (DDD) which is the product of the particle fluence $\Phi$ and the so-called NIEL (non-ionizing energy loss) $dE_{de}/d\chi$ which was calculated by means of the SR (Screened Relativist) treatment \cite{Boschini,Leroy}:
\begin{equation}\label{Eq1}
\frac{dE_{de}}{d \chi }=\frac{N}{A} \int _{E_{d}}^{E_{R}^{max}}E_{R}L \left( E_{R} \right) \frac{d \sigma  \left( E,E_{R} \right) }{dE_{R}}dE_{R}
\end{equation}
where $\chi$  is the absorber thickness in g/cm$^2$, where $N$ is the Avogadro constant; $A$ is the atomic weight of the medium; $E$ is the kinetic energy of the incoming particle; $E_R$ and $E_{Rmax}$ are the recoil kinetic energy and the maximum energy transferred to the recoil nucleus respectively; $E_d$ the displacement threshold energy; $L(E_{R})$ is the Lindhard partition function; $d\sigma(E,E_{R})/dE_{R}$ is the differential cross section for elastic Coulomb scattering for electrons or protons on nuclei. $DDD(E_d)$ also depends on the displacement threshold energy $E_{d}$.
\begin{figure}[h!]
\centering
		\includegraphics[width=0.9\textwidth]{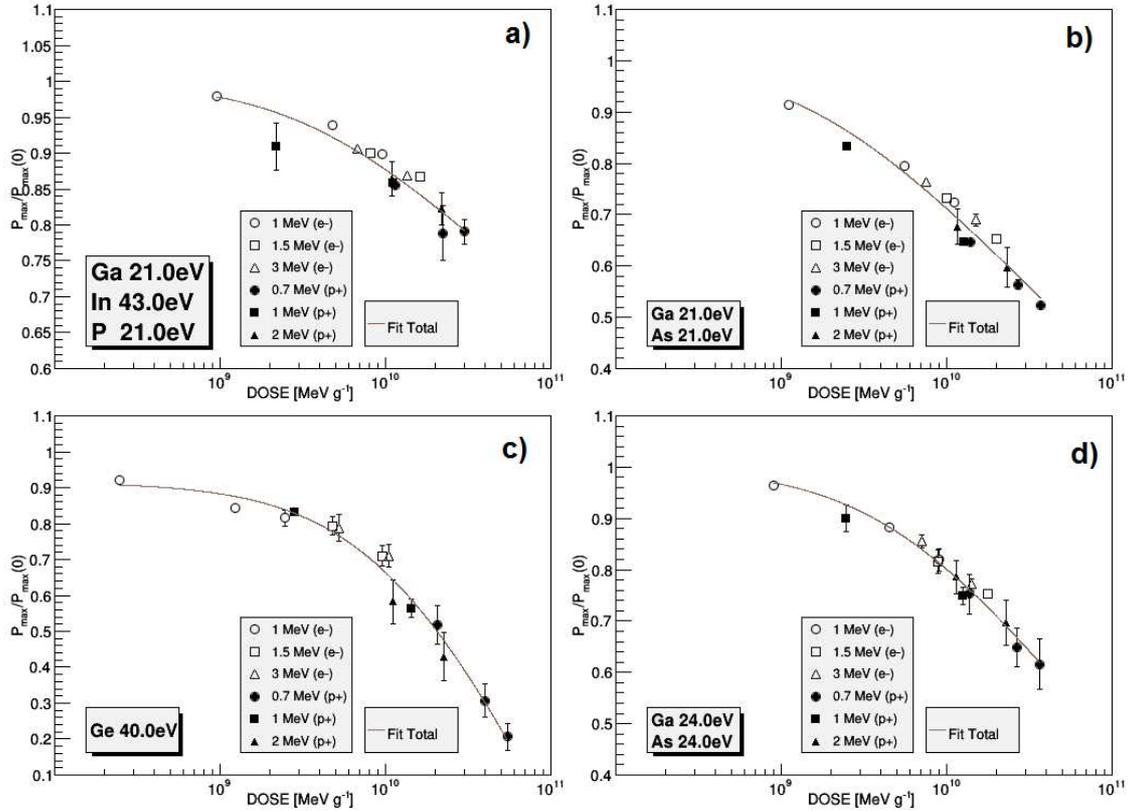}
\caption{Optimal fit of $P_{max}/P_{max}(0)$ as function of the dose for the top cell (a), mid cell (b), bottom cell (c) and TJ solar cell (d).}
\label{fig:3}
\end{figure}
For electrons, there is no relevant kinetic energy variation along the path inside the absorber (i.e. the TJ solar cell). On the contrary, such a change occurs for protons. In fact, their actual energy, in each junction, can be estimated by means of SRIM \cite{Ziegler}. In the current study, the $DDD$ for the TJ solar cells are those evaluated for the middle GaAs cell. The relative degradation of Pmax (Fig. \ref{fig:3}), Isc, and Voc obtained after irradiation for the bottom cell exhibit an expected sudden drop. This effect is related to the specific layout of the Ge component cell and is explained in \cite{Baur}. Therefore, the three sets (Pmax/Pmax(0), Isc/Isc(0), and Voc/Voc(0)) of experimental data were interpolated using the expression:
\begin{equation}\label{Eq2}
\left( 1-C_{1} \right) -C \cdot log_{10} \left[ 1+\frac{D^{NIEL} \left( E_{d} \right) }{D_{x}} \right]
\end{equation}
where $C_1$, $C$ and $D_{x}$ are obtained by a fit to the data in which the NIEL threshold energy, $E_{d}$, was moved to minimize the differences among electron and proton data with respect to the corresponding curve obtained from equation \ref{Eq2}.
It should also be noted that $C_{1}$ is only relevant for the bottom cell. The optimal fits (including the data for Isc and Voc) for GaInP, GaAs and Ge were obtained for $E_d\approx$ 21, 21, 43 and  40 eV for Ga, As, In and Ge respectively; for the TJ cell approximated to a GaAs single cell the displacement threshold found was $\approx$ 24 eV for Ga and As.

\section{DLTS Analysis}
\begin{figure}[h!]
\centering
		\includegraphics[width=0.5\textwidth]{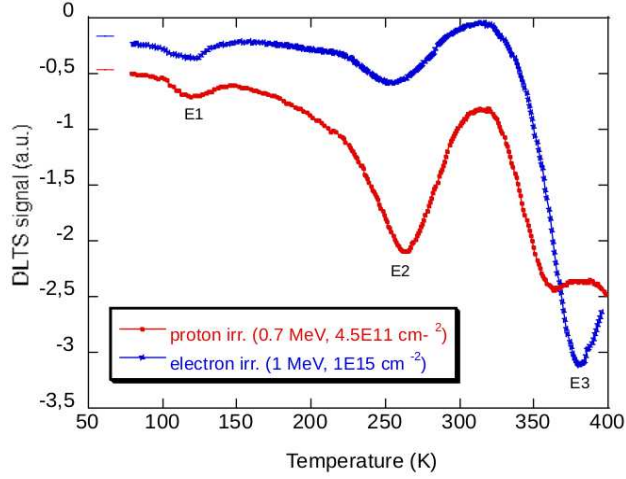}
\caption{Comparison of the DLTS spectra of two middle junctions irradiated by protons and electrons respectively. Emission rate =46 s$^{-1}$, pulse width=500 $\mu$s, reverse voltage Vr=-1.5 V, pulse voltage V1=-0.1 V.}
\label{fig:5}
\end{figure}

\begin{figure}[h!]
\centering
		\includegraphics[width=0.7\textwidth]{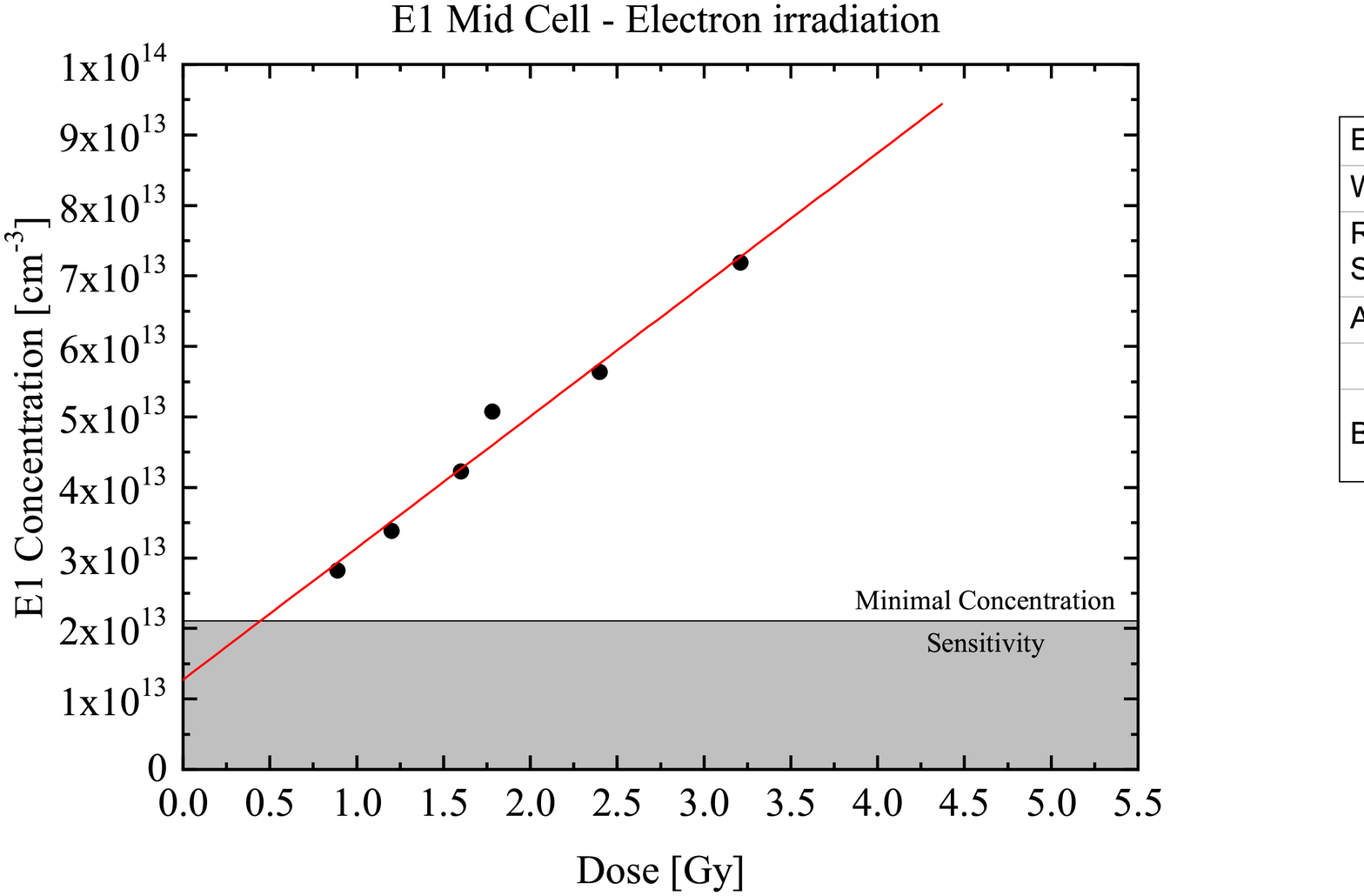}
\caption{Concentration of E1 traps induced by electron irradiation in middle sub cell as a function of Displacement Damage Dose (obtained with Ed = 21 eV).}
\label{fig:6}
\end{figure}

\begin{figure}[h!]
\centering
		\includegraphics[width=0.7\textwidth]{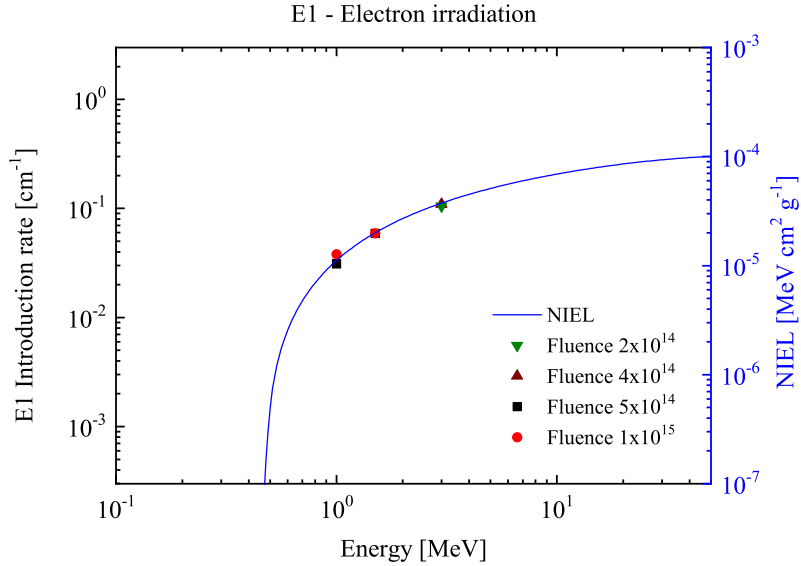}
\caption{E1 traps introduction rate as a function of incoming electron energy: right scale NIEL values in GaAs sub cell for electrons.}
\label{fig:7}
\end{figure}

\begin{figure}[h!]
\centering
		\includegraphics[width=0.7\textwidth]{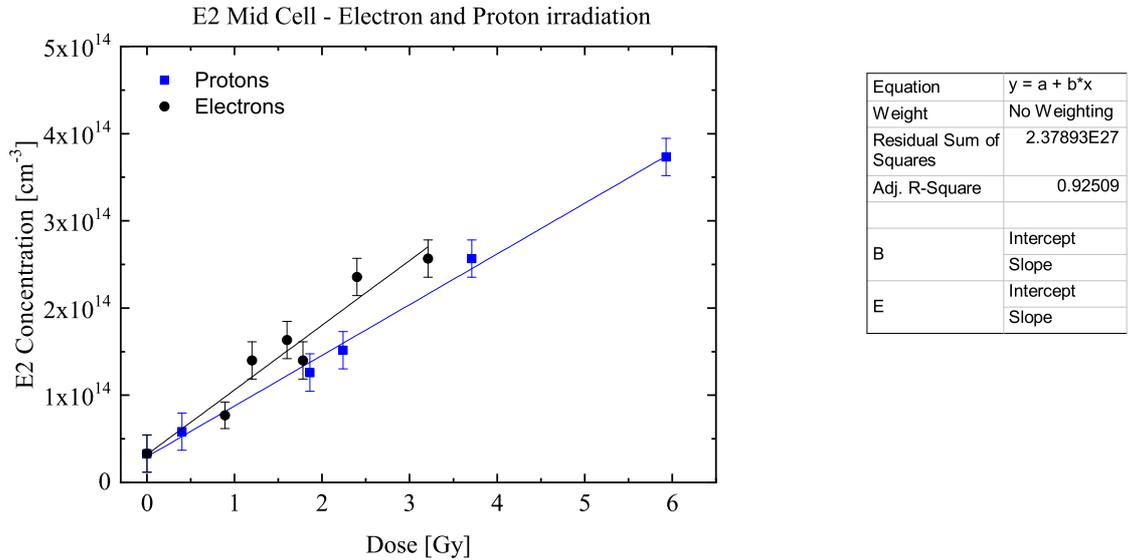}
\caption{Concentration of E2 traps induced by electron and proton irradiation in middle sub cell as a function of Displacement Damage Dose (obtained with $E_{d}$ = 21 eV).}
\label{fig:8}
\end{figure}

\begin{figure}[h!]
\centering
	\includegraphics[width=0.7\textwidth]{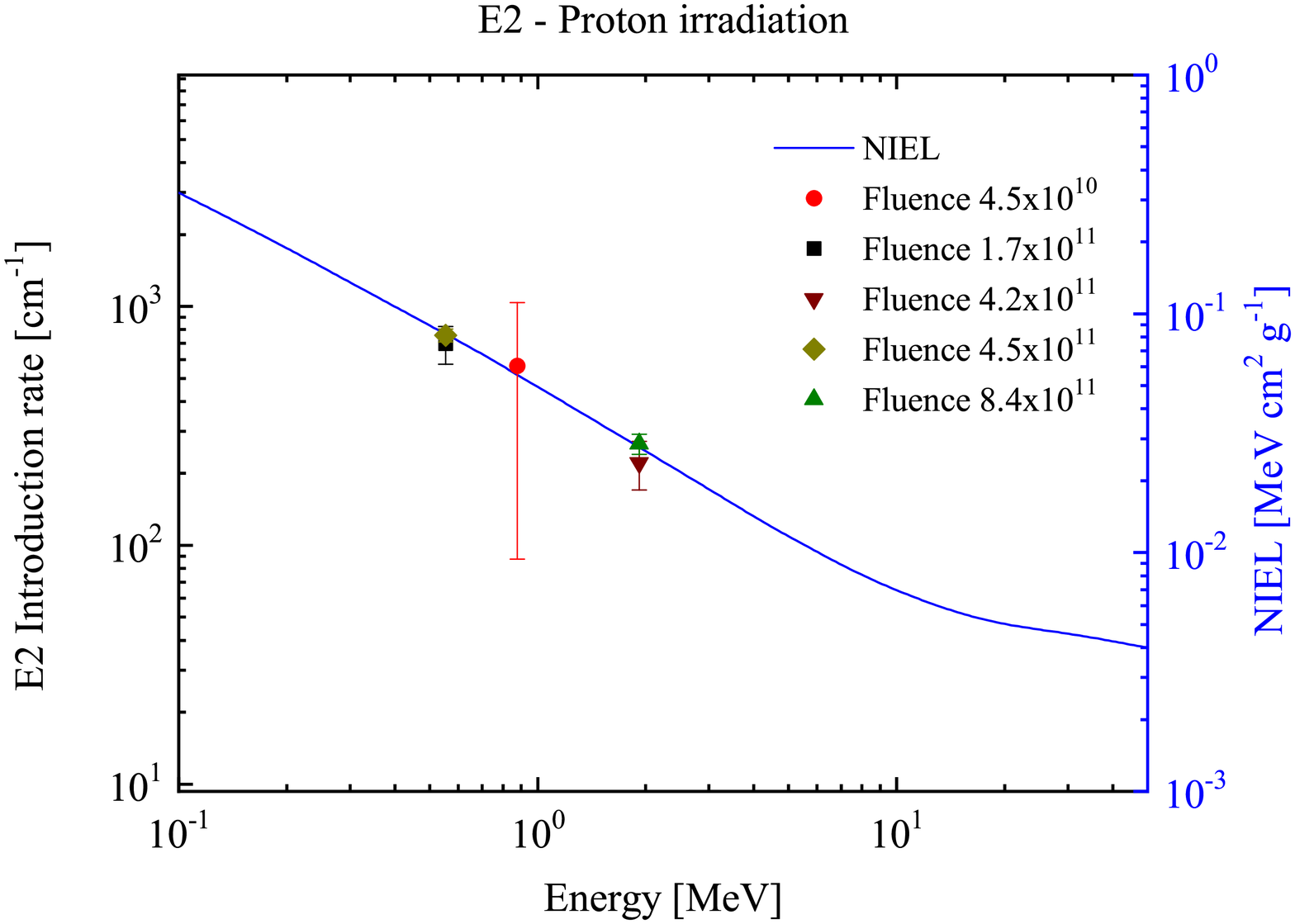}
	\includegraphics[width=0.7\textwidth]{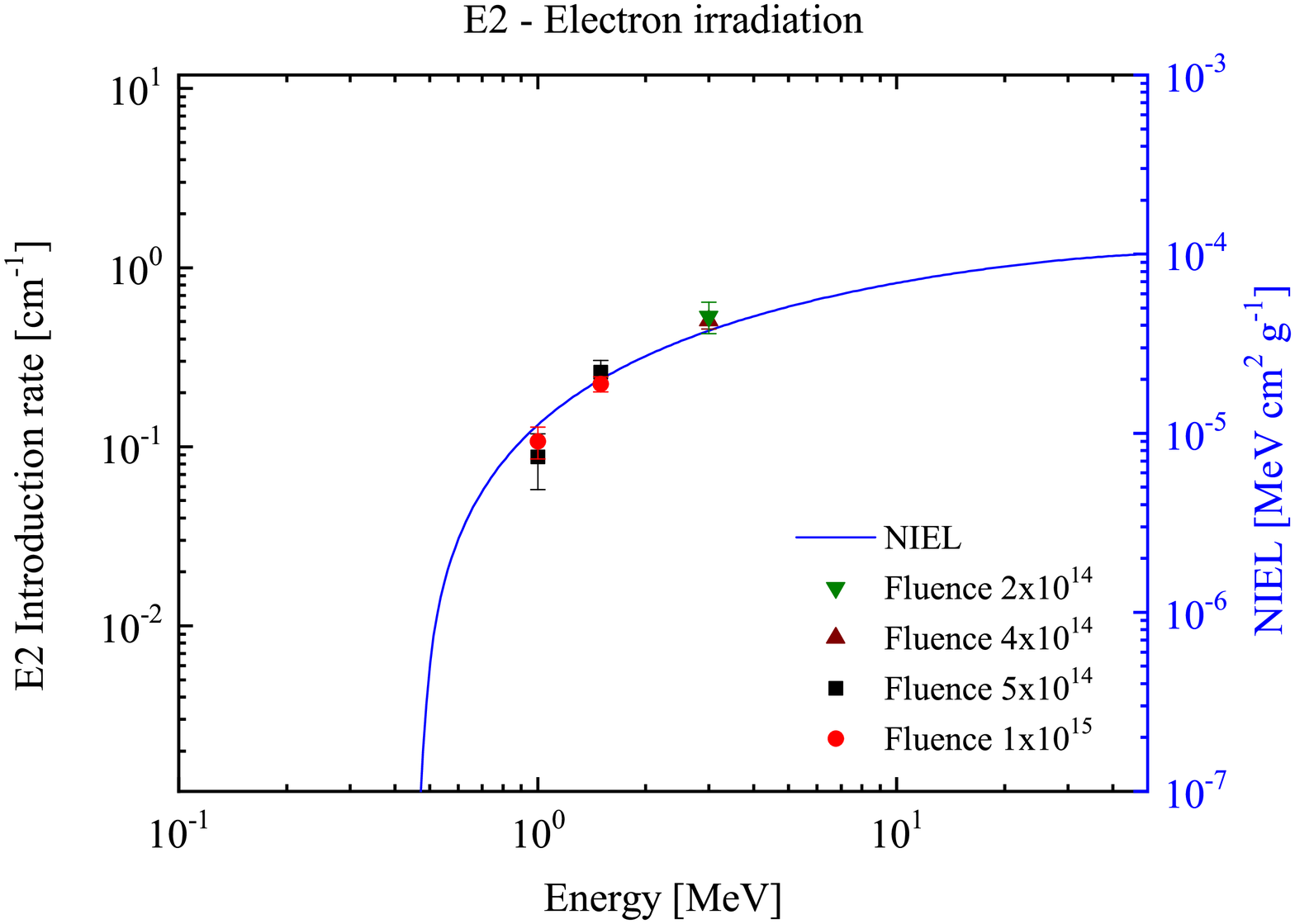}
\caption{E2 traps introduction rate as a function of incoming proton (top) and electron (bottom) energy: right scale NIEL values in GaAs sub cell for protons and electrons.}
\label{fig:9}
\end{figure}

In order to perform DLTS (Deep Level Transient Spectroscopy) investigations of deep levels induced by electron and proton irradiation, mesa-structures of 0.5 mm in diameter have been prepared on top and middle junctions by means of optical photolithography and metal evaporation.
\begin{figure}[h!]
\centering
		\includegraphics[width=0.5\textwidth]{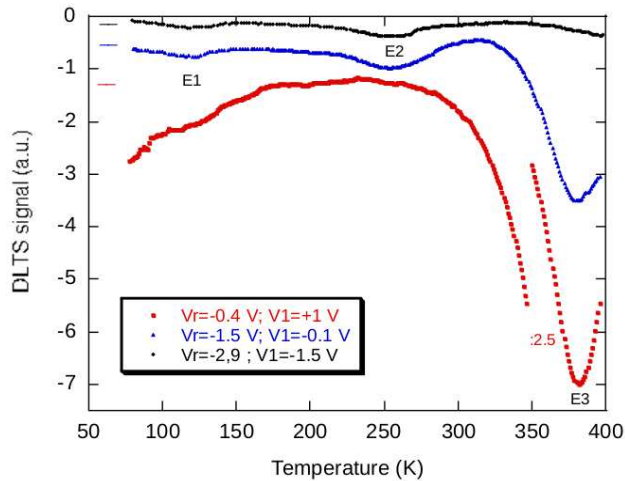}
\caption{DLTS spectra obtained on a middle junction irradiated with electrons (1 MeV at a fluence of $10^{15}$ cm$^{-2}$) using different reverse voltage Vr at fixed pulse amplitude of 1.4 V. Emission rate =46 s$^{-1}$, pulse width=500 ms, V1=pulse voltage.}
\label{fig:4}
\end{figure}
Fig. \ref{fig:4} shows the DLTS spectra of an electron irradiated middle junction obtained using different reverse voltages Vr (-0.4 V, -1.5 V, and -2.9 V) at fixed pulse amplitude (1.4 V).
With increasing reverse voltage Vr the spectra show the characteristics of regions at increasing distance from the n+/p interface. The high temperature peak (activation energy = 0.71 eV) labelled E3, which is also present in the non-irradiated samples,  is likely to correspond to a defect at the junction interface. For higher reverse biases the DLTS spectra show the presence of at least two levels, labelled E1 (0.21 eV) and E2 (0.45 eV). These levels are not observed in non-irradiated samples and show a density which increases with the absorbed dose, hence they are attributed to  irradiation induced defects.
In Fig. \ref{fig:5} the DLTS spectrum of a middle cell diode irradiated with protons  (energy 0.7 MeV and fluence $4.5\times10^{11}$ cm$^{-2}$ corresponding to NIEL dose of $3\times10^{10}$ MeV g$^{-1}$) is compared to that of a middle cell diode irradiated with electrons (energy 1 MeV and fluence $1\times10^{15}$ cm$^{-2}$ corresponding to NIEL dose of $1.1\times10^{10}$ MeV g$^{-1}$).
From the analysis of the graphs the most important observations are: a) two or more DLTS peaks at high temperature are present in the proton irradiated sample, while a single peak E3 is present in the electron irradiated sample; b) in both electron and proton irradiated samples the peaks E1 and E2 and a broad low temperature shoulder of  peak E2 are present; c) the ratio of the peak amplitudes E2/E1 is observed to be much larger for the proton irradiated sample than for the electron irradiated one.
The Concentration of E1 traps was correlated to the Displacement Damage Dose in the middle diode irradiated with electrons. Fig. \ref{fig:6} shows that the additional concentration of E1 traps has  a linear dependence on the NIEL Dose due to electron irradiation, with NIEL dose = NIEL (in units of MeV cm$^2$ / g) $\times$ Particle fluency (in part/cm$^2$).
Considering the introduction rate of E1 traps (in units cm$^{-1}$) at energy E (i.e. the  E1 trap concentration divided by the electron fluence), it is proportional to NIEL (in units of MeV cm$^{2}$/g) at energy E with a proportionality constant given by the slope of the E1 traps vs NIEL Dose.
In case of electrons, this introduction rate to NIEL conversion factor is $3\times10^{3}$ [g MeV$^{-1}$ cm$^{-3}$]. Fig. \ref{fig:7} shows the good agreement between the experimental Introduction rate for E1 with respect to the calculated NIEL for electrons in the middle sub cell.
The concentration of E2 traps was correlated to the Displacement Damage Dose in the middle diode irradiated with electrons and protons. The correlation between E2 density and protons/electrons irradiation dose is reported in Fig. \ref{fig:8}. Both electrons and protons give a linear introduction rate, slightly different for the two particles.
The introduction rate for E2 and the NIEL for electrons and protons are plotted against the particle energy in Fig. \ref{fig:9}. Also in this case there is a direct correlation between E2 Introduction Rate as evaluated from DLTS spectra and the NIEL calculated in the middle sub cell. All the conversion factors are reported in table \ref{Tab1}.
\begin{table}[h!]
\centering
\caption{Conversion factors between Trap Introduction Rate TIR and NIEL.}
\begin{tabular}{|c|c|c|}
\hline
\multicolumn{1}{|c}{\textbf{Traps}} &
\multicolumn{1}{|c}{\textbf{Particle}} &
\multicolumn{1}{|c|}{\textbf{TIR/NIEL(g/MeV/cm$^{3}$)}} \\
\hline
\multicolumn{1}{|c}{E1} &
\multicolumn{1}{|c}{electron} &
\multicolumn{1}{|c|}{$3.0\cdot10^{3}$} \\
\hline
\multicolumn{1}{|c}{E2} &
\multicolumn{1}{|c}{electron} &
\multicolumn{1}{|c|}{$1.2\cdot10^{4}$} \\
\hline
\multicolumn{1}{|c}{E2} &
\multicolumn{1}{|c}{proton} &
\multicolumn{1}{|c|}{$9.3\cdot10^{3}$} \\
\hline
\end{tabular}\label{Tab1}
\end{table}

\section{Conclusion And Future Work}
TJ InGaP/GaAs/Ge solar cells and related component cells were\ irradiated with protons and electrons at different energies.  The data were analysed as a function of Displacement Damage Doses.
DLTS analyses, carried out on middle junctions, indicate that complex defects are likely to be introduced at a different rate for electron and proton irradiations. Two main traps are introduced, E1 having an energy of 0.21 eV and E2 having an energy of 0.45 eV above the top of the valence band. These traps are for majority carriers because the high doping associated with the middle cell junction prevents the detection of minority carrier traps. This is the first time that DLTS analysis is conducted on the real sub cells\ and not on  $``$ad hoc$"$  p/n junction that differs consistently from the real structure of space solar cells.
These set of measurements of Pmax, Isc, Voc and introduction rates of  levels are experimentally supporting the validity of sr-niel treatment for obtaining displacement damage doses.

\vspace*{0.5cm}
\footnotesize{{\bf Acknowledgment:}{\ Part of this work was supported by ESA contract 4000116146/16/NL/HK with title "Non-Ionizing Energy Loss (NIEL) Calculation and Verification".}}

\end{document}